\documentstyle[prl,aps,epsfig]{revtex}
\begin{document}
\draft
\twocolumn[\hsize\textwidth\columnwidth\hsize\csname@twocolumnfalse%
\endcsname

%
\title{
\vspace{0.5cm} $\gamma^* N\rightarrow
\Delta$ transition form factors: a new analysis of the JLab data
on $p(e,e'p)\pi^0$ at Q$^2$=2.8 and 4.0 (GeV/c)$^2$}
\author{Sabit S. Kamalov\cite{Sabit} and Shin Nan Yang}
\address{Department of Physics, National Taiwan University, Taipei
10617, Taiwan}
\author{Dieter Drechsel, Olaf Hanstein and Lothar Tiator}
\address{Institut f\"ur Kernphysik, Universit\"at Mainz, 55099 Mainz, Germany}

\date{\today}
\maketitle

\begin{abstract}
Recent JLab data of the differential cross section for the
reaction $p(e,e'p)\pi^0$ in the  invariant mass region of $1.1 < W
< 1.4$ GeV at four-momentum transfer squared $Q^2=$ 2.8 and 4.0
(GeV/c)$^2$ are  analyzed  with two models, both of which give an
excellent description of most of the existing pion
electroproduction data below $W < 1.5$ GeV. We find that at up to
$Q^2 = 4.0$(GeV/c)$^2$, the extracted helicity amplitudes
$A_{3/2}$ and $A_{/2}$ remain comparable with each other,  implying
that hadronic helicity is not conserved at this range of $Q^2$.
The ratios $E_{1+}/M_{1+}$ obtained show, starting from a small
and negative value at the real photon point, a clear tendency to
cross zero,  and to become positive with increasing $Q^2$. This is
a possible indication of a very slow approach toward the pQCD
region.
Furthermore, we find that the helicity amplitude $A_{1/2}$ and
$S_{1/2}$, but not $A_{3/2}$, starts exhibiting the scaling
behavior at about $Q^2 \ge 2.5$(GeV/c)$^2$.

\end{abstract}
 \vspace{0.5cm}
]

In a recent experiment \cite{Frolov99},  electro-excitation of the
$\Delta$ was studied at $Q^2=$ 2.8 and 4.0 (GeV/c)$^2$ via the
reaction $p(e,e'p)\pi^0$. It was motivated by the possibility of
determining the range of momentum transfers where perturbative QCD
(pQCD) would become applicable. In the limit of $Q^2 \rightarrow
\infty$, pQCD predicts the dominance of helicity-conserving
amplitudes \cite{Brodsky81} and scaling results
\cite{Carlson,Paul}. The hadronic helicity conservation should
have the consequence that the ratio between magnetic dipole
$M_{1+}^{(3/2)}$ and electric quadrupole  $E_{1+}^{(3/2)}$
multipoles, $R_{EM} =E_{1+}^{(3/2)}/M_{1+}^{(3/2)}$, approaches 1.
The scaling behavior predicted by pQCD for the helicity amplitudes
is $A_{1/2}^{\Delta}\sim Q^{-3}$, $A_{3/2}^{\Delta}\sim Q^{-5}$, and
the Coulomb helicity amplitude $S_{1/2}^{\Delta}\sim Q^{-3}$,
resulting in $R_{SM} =S_{1+}^{(3/2)}/M_{1+}^{(3/2)} \rightarrow
const$. On the other hand, in symmetric $SU(6)$ quark models, the
$\gamma N\Delta$ transition can proceed only via the flip of a
single quark spin in the nucleon, leading  to $M_{1+}$ dominance
and $E_{1+} = S_{1+} \equiv 0$. Recent experiments give
nonvanishing ratios  $R_{EM}$ lying between $
-2.5\%$~\cite{Beck97} and $-3.0\%$~\cite{Blanpied97} at $Q^2=0$.
This has been widely taken as an indication of  a deformed
$\Delta$, namely, an admixture of a D state in the $\Delta$.
Accordingly, the question of how $R_{EM}$ would evolve from a very
small negative value at $Q^2 = 0$ to $+100\%$ at sufficiently high
$Q^2$, has attracted  great interest both theoretically and
experimentally.

In Ref. \cite{Frolov99}, the differential cross sections were
measured in the invariant mass region of $\,\,1.1 < W < 1.4$ GeV.
Two methods were used to extract the contributing multipoles. The
first one,  which is model and energy independent, consisted of
making approximate multipole fits to angular distributions
independently at each $W$, assuming $M_{1+}$ dominance, and only
$S$ and $P$ wave contributions \cite{Adler72}. Another extraction
of the resonance amplitudes was performed using the effective
Lagrangian method \cite{Davidson86}. In this model-dependent
analysis, the resonant multipoles are expressed as a sum of
background and resonance amplitudes, both prescribed by an
effective Lagrangian, and unitarized with the K-matrix method. The
parameters in the model were fitted to data points with energy $W$
only up to $1.31$ GeV. The ratios $R_{EM}$ and $R_{SM}$ extracted
with these two methods are both small, negative, and tending to
more negative values with increasing $Q^2$,  indicating that pQCD
is  not yet applicable in this region of $Q^2$. Recently, it was
shown \cite{KY99} that the  $Q^2$-dependence of the ratios
$R_{EM}$ and $R_{SM}$ extracted in Ref. \cite{Frolov99} can be
explained in a dynamical model for electromagnetic production of
pions, together with a simple scaling assumption for the bare
$\gamma^* N \Delta$ form factors.

Because of the significance of the physics involved in the $Q^2$
evolution of $R_{EM}$ and $R_{SM}$, it is important to employ the
best possible extraction method in the analysis of the data.  In
fact, the values of $R_{EM}$ and $R_{SM}$  extracted with the two
methods used in Ref. \cite{Frolov99} differ from each other by
factors of 2 and 1.5 at $Q^2=$ 2.8 and 4.0 (GeV/c)$^2$,
respectively. In this letter, we present the results of a new
analysis of the data of Ref.~\cite{Frolov99}, using a new version
(hereafter called MAID) \cite{MAID00} of the unitary isobar model
developed at Mainz (hereafter called MAID98) \cite{MAID98}, and
the dynamical model developed recently in Ref.~\cite{KY99}, which
both give excellent descriptions of most of the existing pion
photo- and electroproduction data \cite{MAID00}. Our analysis  is
similar to the second method used in \cite{Frolov99} in the sense
that it also makes use of a model. However, we fit all the data
points measured up to $W=$ 1.4 GeV and obtain smaller values of
$\chi^2$ per d.o.f.

 In the dynamical approach to pion photo- and electroproduction
\cite{Tanabe85}, the t-matrix can be expressed as
$t_{\gamma\pi}(E)=v_{\gamma\pi}+v_{\gamma\pi}\,g_0(E)\,t_{\pi
N}(E)\,$ and the physical multipoles in channel $\alpha$ are given
by
\begin{eqnarray}
& &t_{\gamma\pi}^{(\alpha)}(q_E,k)
=\exp{(i\delta^{(\alpha)})}\,\cos{\delta^{(\alpha)}}
\nonumber\\&\times& \left[v_{\gamma\pi}^{(\alpha)}(q_E,k) +
P\int_0^{\infty} dq' \frac{q'^2R_{\pi
N}^{(\alpha)}(q_E,q')\,v_{\gamma\pi}^{(\alpha)}(q',k)}{E-E_{\pi
N}(q')}\right], \label{eq:backgr}
\end{eqnarray}
where $v_{\gamma\pi}$ is the transition potential for
$\gamma^*N \rightarrow \pi N$, and $t_{\pi N}$ and $g_0$ denote
the $\pi N$ t-matrix and free propagator, respectively, with $E
\equiv W$ the total energy in the CM frame.
$\delta^{(\alpha)}$ and $R_{\pi N}^{(\alpha)}$ are the $\pi
N$ scattering phase shift and reaction matrix in channel $\alpha$,
respectively; $q_E$ is the pion on-shell momentum and $k=|{\bf
k}|$ is the photon momentum.

In a resonant channel like (3,3) in which the $\Delta(1232)$ plays
a dominant role, the transition potential $v_{\gamma\pi}$ consists
of two terms, $v_{\gamma\pi}(E)=v_{\gamma\pi}^B +
v_{\gamma\pi}^{\Delta}(E),$ where $v_{\gamma\pi}^B$ is the
background transition potential and $v_{\gamma\pi}^{\Delta}(E)$
corresponds to the contribution of the bare $\Delta$. The
resulting t-matrix can be decomposed into two terms \cite{KY99}
$t_{\gamma\pi}(E)=t_{\gamma\pi}^B + t_{\gamma\pi}^{\Delta}(E)$,
where
$t_{\gamma\pi}^B(E)=v_{\gamma\pi}^B+v_{\gamma\pi}^B\,g_0(E)\,t_{\pi
N}(E), $ and
$t_{\gamma\pi}^\Delta(E)=v_{\gamma\pi}^\Delta+v_{\gamma\pi}^\Delta\,g_0(E)
\,t_{\pi N}(E). $ Here $t_{\gamma\pi}^B$ includes the
contributions from the nonresonant background  and renormalization
of the  vertex $\gamma^*N\Delta$. The advantage of such a
decomposition is that all the processes which start with the
excitation of the bare $\Delta$  are summed up in
$t_{\gamma\pi}^\Delta$. Note that the multipole decomposition of
both $t_{\gamma\pi}^B$ and $t_{\gamma\pi}^\Delta$ would take the
same form as Eq. (\ref{eq:backgr}).

For a correct description of the resonance contributions we need,
first of all, a reliable description of the nonresonant part of
the amplitude. In MAID98, the background contribution was
described by Born terms obtained  with an energy dependent mixing
of pseudovector-pseudoscalar $\pi NN$ coupling and t-channel
vector meson exchanges, namely, $t_{\gamma\pi}^{B,\alpha}({\rm
MAID}98)=v_{\gamma\pi}^{B,\alpha}(W,Q^2)$. The mixing parameters
and coupling constants were determined from an analysis of
nonresonant multipoles in the appropriate energy regions. In the
new version of MAID, the $S$, $P$, $D$ and $F$ waves of the
background contributions are complex numbers defined in accordance
with the K-matrix approximation,
\begin{equation}
 t_{\gamma\pi}^{B,\alpha}({\rm MAID})=
 \exp{(i\delta^{(\alpha)})}\,\cos{\delta^{(\alpha)}}
 v_{\gamma\pi}^{B,\alpha}(W,Q^2).
\label{eq:bg00}
\end{equation}
From Eqs. (\ref{eq:backgr}) and  (\ref{eq:bg00}), one finds that
the difference between the background terms of MAID and of the
dynamical model is that off-shell rescattering contributions
(principal value integral) are not included in MAID. To take
account of the inelastic effects at the higher energies, we
replace $\exp{i(\delta^{(\alpha)})} \cos{\delta^{(\alpha)}} =
\frac 12 [\exp{(2i\delta^{(\alpha)})} +1]$ in Eqs.
(\ref{eq:backgr}) and (\ref{eq:bg00}) by $\frac 12
[\eta_{\alpha}\exp{(2i\delta^{(\alpha)})} +1]$, where
$\eta_{\alpha}$ is the inelasticity. In our actual calculations,
both the $\pi N$ phase shifts $\delta^{(\alpha)}$ and
 inelasticity parameters $\eta_{\alpha}$ are taken from the analysis
of the GWU group \cite{VPI97}. Furthermore, the off-shell
rescattering effects in the dynamical model are evaluated with the
reaction matrix $R_{\pi N}^{(\alpha)}(q_E,q')$ as prescribed by a
meson exchange model \cite{Hung94}.

Following  Ref.~\cite{MAID98},  we assume a Breit-Wigner form
for the resonance contribution
${\cal A}^{R}_{\alpha}(W,Q^2)$ to the total multipole amplitude,
\begin{equation}
{\cal A}_{\alpha}^R(W,Q^2)\,=\,{\bar{\cal A}}_{\alpha}^R(Q^2)\,
\frac{f_{\gamma R}(W)\Gamma_R\,M_R\,f_{\pi
R}(W)}{M_R^2-W^2-iM_R\Gamma_R} \,e^{i\phi}, \label{eq:BW}
\end{equation}
where $f_{\pi R}$ is the usual Breit-Wigner factor describing the
decay of a resonance $R$ with total width $\Gamma_{R}(W)$ and
physical mass $M_R$. The expressions for $f_{\gamma R}, \, f_{\pi
R}$ and $\Gamma_R$ are given in Ref.~\cite{MAID98}. The phase
$\phi(W)$ in Eq. (\ref{eq:BW}) is introduced to adjust the phase
of the total multipole to  equal  the corresponding $\pi N$  phase
shift $\delta^{(\alpha)}$. Because  $\phi=0$ at resonance,
$W=M_R$, this phase does not affect the $Q^2$ dependence of the
$\gamma N R$ vertex.

We now concentrate on the $\Delta(1232)$. In this case the
magnetic dipole $({\bar{\cal A}}_M^{\Delta})$ and the electric
quadrupole $({\bar{\cal A}}_E^{\Delta})$ form factors are related
to the conventional electromagnetic helicity amplitudes
$A^\Delta_{1/2}$, $A^\Delta_{3/2}$ and $S^\Delta_{1/2}$ by
\begin{eqnarray}
{\bar{\cal A}_M^\Delta}(Q^2)&=&-\frac{1}{2}(A^\Delta_{1/2} +
\sqrt{3} A^\Delta_{3/2}),\,\\
 {\bar{\cal A}_E^\Delta}(Q^2)&=&\frac{1}{2}
(-A^\Delta_{1/2} + \frac{1}{\sqrt{3}} A^\Delta_{3/2}),\,\\
 {\bar{\cal A}_S^\Delta}(Q^2))&=&-\frac{1}{\sqrt{2}}S^\Delta_{1/2},\,
\end{eqnarray}
where $k^2=Q^2+[(W^2-m_N^2-Q^2)/2W]^2$. We stress
that the physical meaning of these resonant amplitudes in
different models is different\cite{KY99,Sato}. In MAID, they
contain contributions from the background excitation and describe
the so called "dressed" $\gamma N\Delta$ vertex. However, in the
dynamical model the background excitation is included in ${\cal
A}^{B}_{\alpha}$ and the electromagnetic vertex ${\bar{\cal
A}}_{\alpha}^\Delta(Q^2)$ corresponds to the
 "bare" vertex.

In the dynamical model of Ref. \cite{KY99}, a scaling assumption
was made concerning the (bare) form factors ${\bar{\cal
A}}_{\alpha}^\Delta(Q^2)$, namely, that all of them have the same
$Q^2$ dependence. In the present analysis, we do not impose the
scaling assumption and write, for electric ($\alpha=E$), magnetic
($\alpha=M$) and Coulomb ($\alpha=S$) multipoles,
\begin{eqnarray}
{\bar{\cal A}}_{\alpha}^{\Delta}(Q^2)=X_{\alpha}^{\Delta}(Q^2)\,{\bar{\cal
A}}_{\alpha}^{\Delta}(0) \frac{ k}{k_W}\,F(Q^2),
\end{eqnarray}
where  $k_W = (W^2 - m_N^2)/2W$. The form factor $F$ is taken to
be $ F(Q^2)=(1+\beta\,Q^2)\,e^{-\gamma Q^2}\,G_D(Q^2),$ where
$G_D(Q^2)=1/(1+Q^2/0.71)^2$ is the usual dipole form factor. The
parameters $\beta$ and $\gamma$ were determined by setting
$X_M^{\Delta}=1$ and fitting ${\bar{\cal A}}_{M}^{\Delta}(Q^2)$ to
the data for $G_M^*$ as defined in \cite{KY99,MAID98,Ash}.
%
The values of ${\bar{\cal A}}_M^{\Delta}(0)$ and ${\bar{\cal
A}}_E^{\Delta}(0)$ were determined by fitting to the multipoles
obtained in the recent analyses of the Mainz \cite{HDT} and GWU
\cite{VPI97} groups. Both $X_E$ and $X_S$ are to be determined by
the experiment with $X^\Delta_{\alpha}(0)=1$. Note that deviations
from $X^\Delta_{\alpha}=1$ value will indicate a violation of the
scaling law. Similar treatment is also applied to the $N^*(1440)$
resonance with two additional parameters $X_M^{P11}$ and
$X_S^{P11}$ corresponding to the transverse and longitudinal
resonance transitions in the isospin 1/2 channel.
\begin{figure}[h]
\begin{center}
\epsfig{file=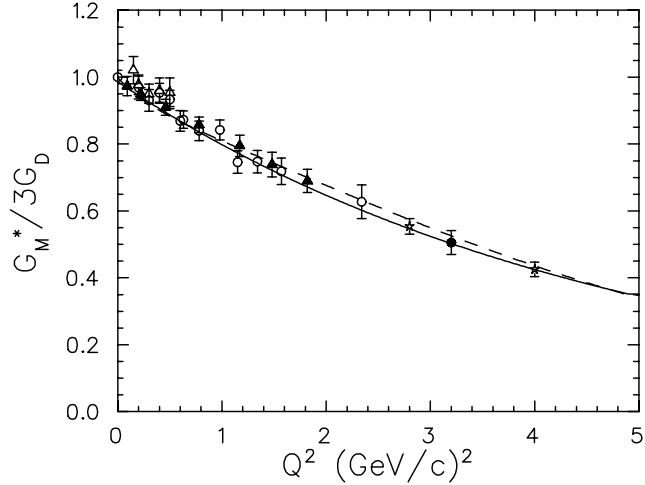, width=3.3in}
\end{center}
\caption{ The $Q^2$ dependence of the $G_M^*$ form factor. The
solid and dashed curves are the results of the MAID and dynamical
model analyses, respectively. The data at $Q^2$=2.8 and 4.0
$(GeV/c)^2$ are from Ref.\protect\cite{Frolov99}, other data from
Refs.\protect\cite{EXPM}. } \label{fig1}
\end{figure}
\begin{center}
\begin{minipage}{3.4in}
\begin{table}[h]
\caption{\small Our results for the ratios $R_{EM}$ and $R_{SM}$,
at $Q^2$=2.8 (upper row) and 4.0 (lower row) (GeV/c)$^2$,
extracted from a global fit to the data with MAID and the
dynamical model as discussed in the text. Results from Ref.
\protect\cite{Frolov99} are listed for comparison. Ratios are
given in (\%).}
\begin{tabular}{c|ccc}
 models        &  MAID  &  DM  & Ref. \cite{Frolov99} \\
\hline $R_{EM}^{(p\pi^0)}$  & $ -0.56\pm 0.33 $ & $-1.28\pm 0.32$
& $-2.00\pm 1.7$\\
                     & $  0.09\pm 0.50 $ & $-0.84\pm 0.46$ & $-3.1\pm  1.7$ \\
\hline $R_{SM}^{(p\pi^0)}$  & $ -9.14\pm 0.54 $ & $-11.65\pm 0.52$
& $-11.2\pm 2.3$
\\
                     & $-13.37\pm 0.95 $ & $-17.70\pm 1.0 $ & $-14.8\pm 2.3 $ \\
\hline $ G_M^*\times 100$   & $  6.78\pm 0.05 $ & $ 7.00\pm 0.04$
& $ 6.9\pm 0.4$ \\
       $\times 100$  &  $  2.86\pm 0.02 $ & $ 3.04\pm 0.02$
& $ 2.9\pm 0.2$\\ \hline $\chi^2$             &     1.02
&     1.46  & 1.60 \\
                     &      1.14          &     1.28  & 1.45 \\
\end{tabular}
\end{table}
\end{minipage}
\end{center}

The dynamical model and MAID are used to analyze the recent JLab
differential cross section data on $p(e,e'p)\pi^0$ at high $Q^2$.
All measured data, 751 points at $Q^2$=2.8 and 867 points at
$Q^2$=4.0 (GeV/c)$^2$ covering the entire energy range $1.1 < W <
1.4$ GeV, are included in our global fitting procedure.
We obtain a very good fit to the measured differential cross
sections. In fact, the values of $\chi^2/d.o.f.$ for our two
models are smaller than those obtained in Ref. \cite{Frolov99}
(see Table 1). Our results for the $G_M^*$ form factor are shown
in Fig. 1. Here the best fit is obtained with $\gamma=0.21$
(GeV/c)$^{-2}$ and $\beta=0$ in the case of MAID, and
$\gamma=0.40$ (GeV/c)$^{-2}$ and $\beta=0.52$(GeV/c)$^{-2}$ in the
case of the dynamical model.
\begin{figure}[h]
\begin{center}
\epsfig{file=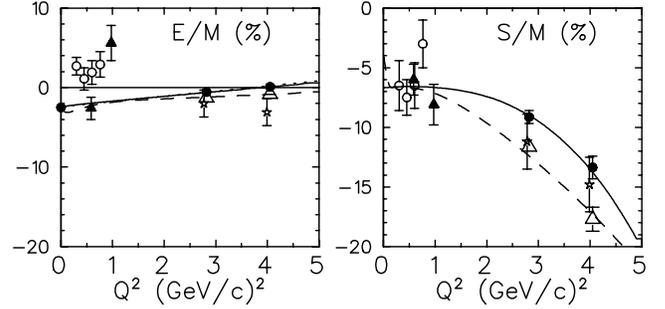,width=3.3in}
\end{center}
\caption{ The $Q^2$ dependence of the ratios $R_{EM}^{(p\pi^0)}$
and $R_{SM}^{(p\pi^0)}$ at $W=1232$ MeV. The solid and dashed
curves are the MAID and dynamical model results, respectively,
obtained with a violation of the scaling assumption. Results of
previous data analysis at $Q^2=0$ from Ref.\protect\cite{Beck97},
data at $Q^2$=2.8 and 4.0 $(GeV/c)^2$ from
Ref.\protect\cite{Frolov99} (stars). Results of our analysis at
$Q^2$=2.8 and 4.0 $(GeV/c)^2$  are obtained using MAID ($\bullet$)
and the dynamical models ($\bigtriangleup$). Other data from
Ref.\protect\cite{EXPR}.}
\end{figure}

With the resonance parameters $X_{\alpha}^{\Delta}(Q^2)$
determined from the fit, the ratios $R_{EM}=Im E_{1+}/Im M_{1+}$
and $ R_{SM}=Im S_{1+}/Im M_{1+}$ of the total multipoles and the
helicity amplitudes $A_{1/2}$ and $A_{3/2}$ can then be calculated
at resonance. We perform the calculations for both physical
$(p\pi^0)$ and isospin 3/2 channels and find them to agree with
each other. The extracted $Q^2$ dependence of the
$X_{\alpha}^{\Delta}$ parameters is: $X_{E}^{\Delta}({\rm
MAID})=1-Q^2/3.7\,, X_{E}^{\Delta}({\rm DM})=1+Q^4/2.4$, $
X_{S}^{\Delta}({\rm MAID})=1+Q^6/61\,, X_{S}^{\Delta}({\rm
DM})=1-10Q^2$, with $Q^2$ in units (GeV/c)$^2$.

Our extracted values for $R_{EM}$ and $R_{SM}$ and a comparison
with the results of Ref.~\cite{Frolov99} are presented in Table 1
and shown in Fig. 2. The main difference between our results and
those of Ref.~\cite{Frolov99} is that our values of $R_{EM}$ show
a clear tendency to cross zero and change sign as $Q^2$ increases.
This is in contrast with the results obtained in the original
analysis \cite{Frolov99} of the data which concluded that $R_{EM}$
would stay negative and tend toward more negative values with
increasing $Q^2$. Furthermore, we find that the absolute value of
$R_{SM}$ is strongly increasing. In terms of helicity amplitudes,
our results for a small $R_{EM}$ can be understood in that the extracted
$A_{3/2}$ remains as large as the helicity conserving $A_{1/2}$  up to
$Q^2 = 4.0 (GeV/c)^2$, resulting in a small $E_{1^+}$.

Finally, we show our results for $Q^3A_{1/2}^{\Delta},
Q^5A_{3/2}^{\Delta},$ and $Q^3S_{1/2}^{\Delta}$ in Fig. 3. The bare form
factors obtained with DM is used since the scaling behavior predicted by
pQCD arises from the $3q$ Fock states in the nucleon and $\Delta$. It is
interesting to see that $S_{1/2}^{\Delta}$ and $A_{1/2}^{\Delta}$
clearly starts exhibiting the pQCD scaling behavior at about $Q^2
\ge 2.5 (GeV/c)^2$. The maximal value for the $Q^3
A_{1/2}^{\Delta}$ which we obtained in this region is about -0.11
GeV$^{5/2}$. This is in between of the asymptotic values for this
quantity discussed in Ref.\cite{Carlson}, i.e., $Q^3
A_{1/2}^{\Delta}=$ -0.08 GeV$^{5/2}$ and -0.17 GeV$^{5/2}$.
However, it is difficult to draw any definite conclusion for
$Q^5A_{3/2}^{\Delta}$. From these results, it appears likely
 that scaling will set in earlier than the helicity
conservation as predicted by pQCD.

\begin{figure}[h]
\begin{center}
\epsfig{file=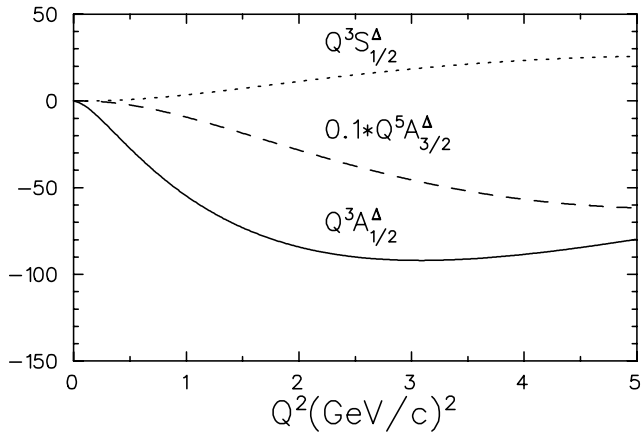,width=3.3in}
\end{center}
\caption{ The $Q^2$ dependence of the  $Q^3 A_{1/2}^{\Delta}$
(solid curve) $Q^5 A_{3/2}^{\Delta}$ (dashed curve) and $Q^3
S_{1/2}^{\Delta}$ (dotted curve) amplitudes (in units 10$^{-3}$
GeV$^{n/2}$) obtained with DM.} \label{fig3}
\end{figure}

In summary, we have re-analyzed the recent JLab data for
electroproduction of the $\Delta(1232)$ resonance via
$p(e,e'p)\pi^0$ with two models for pion electroproduction, both
of which give excellent descriptions of the existing data. We find
that $A_{3/2}^{\Delta}$ is still as large as $A_{1/2}^{\Delta}$ at
$Q^2=4$ (GeV/c)$^2$, which implies that hadronic helicity
conservation is not yet observed in this region of $Q^2$.
Accordingly, our extracted values for $R_{EM}$ are still far from
the pQCD predicted value of $+100\%$. However, in contrast to
previous results we find that $R_{EM}$, starting from a small and
negative value at the real photon point, actually exhibits a clear
tendency to cross zero and change sign as $Q^2$ increases, while
the absolute value of $R_{SM}$ is strongly increasing. In regard
to the scaling, our analysis indicates that $S_{1/2}^{\Delta}$ and
$A_{1/2}^{\Delta}$, but not $A_{3/2}^{\Delta}$, starts exhibiting
the pQCD scaling behavior at about $Q^2 \ge 2.5 (GeV/c)^2$. It
appears likely that the onset of scaling behavior might take place
at a lower momentum transfer than that of hadron helicity
conservation.

It will be most interesting to have data at yet higher momentum
transfer in order to see the region where the helicity amplitude
$A_{1/2}^{\Delta}$ finally dominates over $A_{3/2}^{\Delta}$. It
is only there that we could expect to see  the onset of the
asymptotic behavior of $R_{EM}\rightarrow +100\%$ and
$R_{SM}\rightarrow const$.

\acknowledgements

We are grateful to Paul Stoler and Rick Davidson for useful
communications. S.S.K. would like to thank the Department of
Physics at National Taiwan University for warm hospitality and
gratefully acknowledges the financial support of the National
Science Council of ROC. This work was supported in part by  NSC
under Grant No. NSC89-2112-M002-038, by Deutsche
Forschungsgemeinschaft (SFB443) and by a joint project NSC/DFG
TAI-113/10/0.

\end{document}